\documentclass[%
print,
superscriptaddress,
amsmath,amssymb,amsfonts,
aps,
prl,
floatfix,notitlepage,latexsym
]{revtex4-1}

\usepackage[dvipdfmx]{graphicx}
\usepackage{dcolumn}
\usepackage{bm}
\usepackage{braket}
\usepackage{mediabb} 
\usepackage[dvipdfmx]{attachfile2}

\begin{abstract}
Magnetic oscillation is a generic property of electronic conductors under magnetic fields and widely appreciated as a useful probe of their electronic band structure, i.e., the Fermi surface geometry. However, the usage of the strong static magnetic field makes the measurement insensitive to the magnetic order of the target material. That is, the magnetic order is anyhow turned into a forced ferromagnetic one.
Here we theoretically propose an experimental method of measuring the magnetic oscillation in a magnetic-order-resolved way by using the azimuthal cylindrical vector (CV) beam, an example of topological lightwaves. The azimuthal CV beam is unique in that when focused tightly, it develops a pure longitudinal magnetic field.  We argue that this characteristic focusing property and the discrepancy in the relaxation timescale between conduction electrons and localized magnetic moments allow us to develop the nonequilibrium analog of the magnetic oscillation measurement. Our optical method would be also applicable to metals the under ultra-high pressure of diamond anvil cells.  
\end{abstract}

\begin{document}


\title{Nonequilibrium Magnetic Oscillation with Cylindrical Vector Beams}
\author{Hiroyuki Fujita}
\thanks{Corresponding author}
\affiliation{Institute for Solid State Physics, University of Tokyo, Kashiwa 277-8581, Japan}
\email{h-fujita@issp.u-tokyo.ac.jp}
\author{Masahiro Sato}
\affiliation{Department of Physics, Ibaraki University, 
Mito, Ibaraki 310-8512, Japan}
\email{masahiro.sato.phys@vc.ibaraki.ac.jp}
\date{\today}

\maketitle

\section{Introduction}
When an electric conductor is under a strong magnetic field, the electronic band structure is reconstructed to be Landau levels, and the isoenergy surface of electrons in the momentum space reduce into the so-called Landau tubes. For a fixed Fermi energy, these Landau tubes and their field dependence cause oscillations of various electronic properties as a function of the external magnetic field~\cite{ashcroft1976solid,kittel2004introduction}. If we focus on the electronic conductivity, the oscillation is called the Shubnikov-de Haas effect and if we on the spin polarization (or magnetic susceptibility), that is called the de Haas-van Alphen effect. In the latter case, the oscillation frequency is determined by the area of the extremal orbit on the Fermi surface. Hence, by measuring the oscillation while changing the direction and the strength of the magnetic field, we can know the Fermi surface in the full Brillouin zone.

Although developed decades ago, the magnetic oscillation is still an vital tool in condensed matter physics~\cite{Li1208,Tan287,PhysRevLett.114.176601,Li:2015aa,Chen:2015aa,PhysRevLett.115.057202,Ben-Shalom:2015aa,PhysRevLett.116.046403,PhysRevB.94.104514,Wu:2017aa}. For example, recent discovery of the bulk-like quantum oscillation in a Kondo insulator SmB${}_6$~\cite{Tan287} triggered studies~\cite{Park6599,PhysRevLett.116.246403,PhysRevB.94.165154} searching for the novel charge-neutral Fermi surface~\cite{Coleman:1993aa,mix-valence,PhysRevLett.118.096604,PhysRevLett.119.057603}.

\begin{figure}[htbp]
\centering
\includegraphics[width = 140mm]{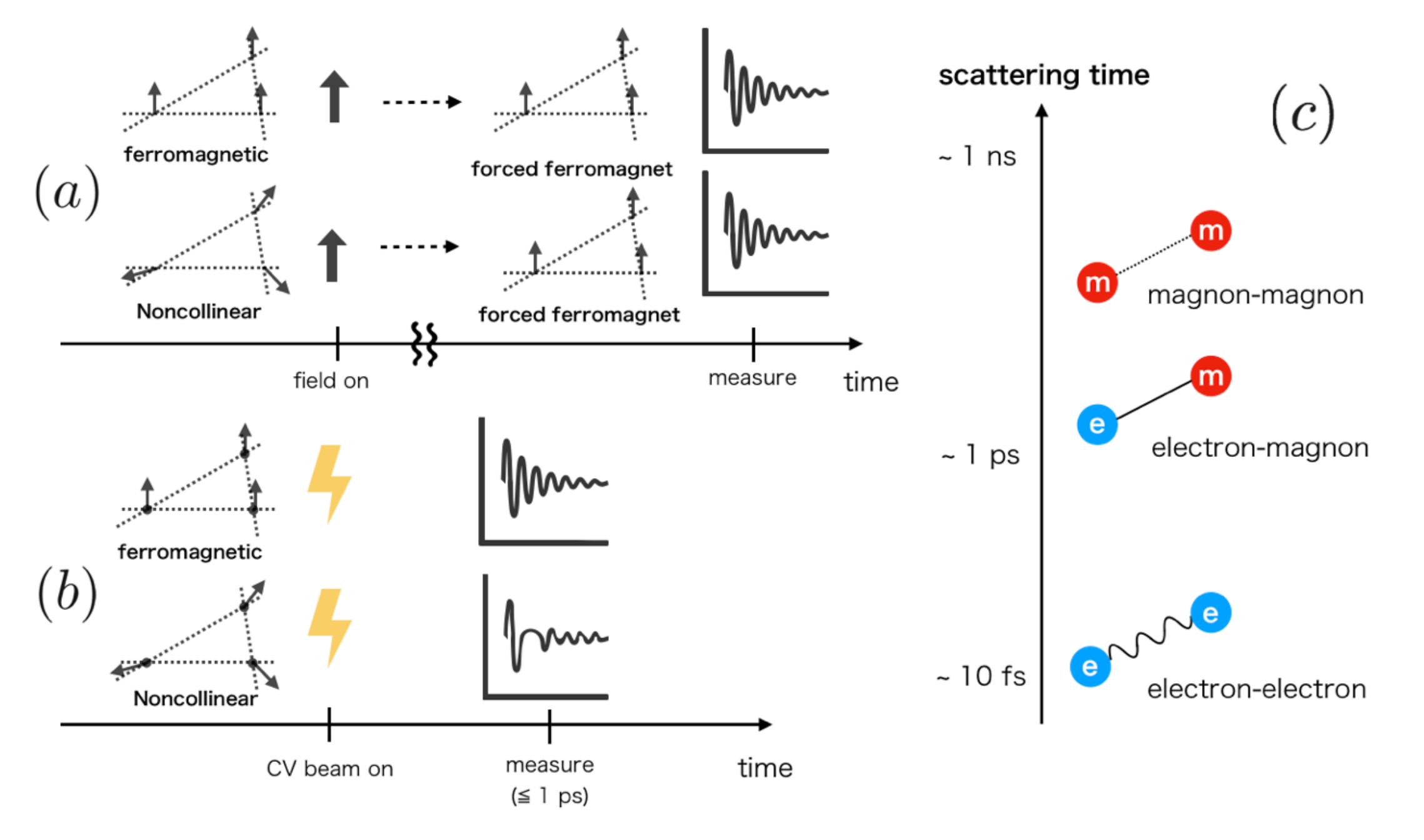}
\caption{$(a)$: Schematics of conventional magnetic oscillation measurement performed to a conducting magnet. Arrows on vertices of the lattice represent localized moments and the thick black arrows do the static magnetic field perpendicular to the lattice. Irrespective to the original spin texture, the static field makes the system to be a forced ferromagnet. $(b)$: Schematics of laser-based nonequilibrium measurement proposed in this paper. We apply a cylindrical vector (CV) beam (explained later) pulse to form the Landau levels to which electrons relax. The latter method has the sensitivity to the magnetic structure of the initial state. Panel $(c)$ shows the hierarchy of the scattering times in solids~\cite{PhysRevLett.41.805,RevModPhys.82.2731,PhysRevB.73.144424}.}
       \label{schem}
  \end{figure}

When the target system is a non-magnetic conductor, the magnetic oscillation is indeed a powerful tool. 
However, if we are to apply that to conducting magnets with localized magnetic moments, a problem arises. 
In this case, the strong static magnetic field applied to measure the oscillation itself affects the magnetic structure. Therefore even if we are interested in the electronic structure of antiferromagnetic, ferrimagnetic, or some non-collinear magnetic states, there is a high possibility that a forced ferromagnetic state [Fig.~\ref{schem}$(a)$] is created by the applied magnetic field. 

Even for the collinear magnetic states, the use of the magnetic oscillation is not effective. Under the strong magnetic field, the collinear magnetic moments point to the direction of the external field. As a result, we cannot obtain the full profile of the Fermi surface because the magnetic oscillation only gives the information about the Fermi surface cross section perpendicular to the applied magnetic field. 

Due to this, the magnetic oscillation is not suitable for the study of the field-induced phase transitions where the strength or direction of the magnetic field have drastic impact on the electronic structure, though there are a number of such materials of our interest. A notable example is pyrochlore iridates where various electronic structures (metal, topological semimetal, and insulator) appear depending on the magnetic structure determined by the direction of the external magnetic field~\cite{Tian:2015aa,Ueda:2017aa}. Another example of the field-induced transition is in thin films of chiral magnets where we observe helical magnetic phase, skyrmion crystal phase, and ferromagnetic state as depending on the strength of the external magnetic field applied vertically to the film~\cite{Muhlbauer:2009aa,Yu:2010aa,Seki_BOOK}. In this paper, we propose an extension of the magnetic oscillation measurement applicable to those field-sensitive materials.

The key idea is to exploit the difference in the relaxation timescale between the conduction electrons and localized spins~\cite{PhysRevLett.41.805,RevModPhys.82.2731,PhysRevB.73.144424}. The energy scale of conduction electrons is of the order of electron volt (eV) while that of spin system is at most of meV. As a result, even though the electrons reach their thermal equilibrium state within 50-500 femtoseconds (fs) after the excitation, it takes more than 1 or 10 picoseconds (ps) for spins to follow that change.
 
Therefore, if we apply a magnetic field pulse of far-infrared or THz frequency (whose timescale is of the order of 100 fs to 1 ps), as Fig.~\ref{schem}$(b)$ shows, conduction electrons would follow the change of the magnetic field through the rapid equilibration to the Landau levels before localized spins respond to that [see also Fig.~\ref{schem}$(c)$].  Then if we measure the conducting-electron spin polarization (or the total magnetization) of this nonequilibrium state, we will see the magnetic oscillation of the electrons under the influence of the initial magnetic structure.

The problem is that if we use the conventional Gaussian laser pulses to apply the magnetic field, there inevitably accompanies the electric field which strongly excites and heats the electron system, smearing out the Landau tube structure we are interested in. Therefore, what we need is thus a source of a ``pure magnetic field" without accompanying the electric field in the optical frequency regime. Equipments used for NMR or ESR experiments cannot generate such a high-frequency magnetic field. As we see below, what is called the azimuthal cylindrical vector (CV) beam meets the above criterion.

\begin{figure}[htbp]
\centering
\includegraphics[width = 150mm]{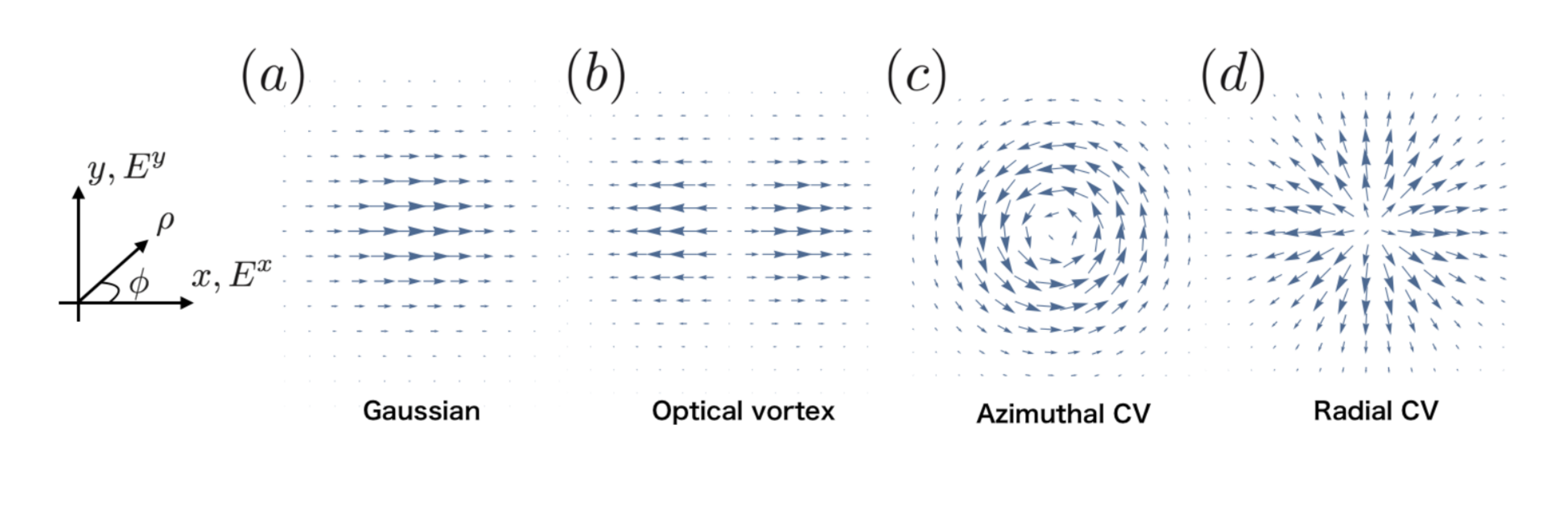}
       \caption{Snapshots of the in-plane ($x$ and $y$) components of the electric field of $(a)$ linearly polarized Gaussian beam, $(b)$ linearly polarized Optical vortex with unit orbital angular momentum, $(c)$ azimuthal CV beam, and $(d)$ radial CV beam. The field amplitude of the Gaussian beam is the strongest at the center while beams $(b)-(d)$ have vanishing in-plane components there due to their topological nature. The size of the arrows in the figure corresponds to the laser amplitude at each point. We show the definition of $(\rho, \phi)$ in the cylindrical coordinate for later use.}
       \label{pol}
  \end{figure}  

\section{Vector beam} 
In this section, we review the properties of the CV beam and how can they be exploited to the observation of the magnetic oscillations.
In modern optics, a class of laser beams called the topological beams such as optical vortices~\cite{PhysRevA.45.8185,9780511795213} and CV beams~\cite{Youngworth:00,Zhan:02,Zhan:09} are intensively studied experimentally. These beams are characterized with the topologically nontrivial spatial profile originating from either the spiral phase structure (optical vortices) or vortices in the polarization vector (CV beams) and can be generated by using holograms, structured phase plates, and so on~\cite{9780511795213,Zhan:09}. Mathematically, CV beam can be seen as a superposition of optical vortices.

 In Fig.~\ref{pol}, we compare the conventional Gaussian beam with those topological beams. The arrows in the figure correspond to the in-plane components of the electric field of those beams propagating perpendicularly 
to the $x$-$y$ plane. We see that topological beams have vanishing in-plane components at the center reflecting their spatial profile while those of the Gaussians beam are the strongest there. Optical vortices carry an orbital angular momentum (not a spin angular momentum) and their applications 
to condensed matter physics have begun to be actively explored in very recent years~\cite{Hamazaki:10,Terhalle:11,Toyoda:2012aa,Takahashi:13,PhysRevB.93.045205,Fujita2016,PhysRevB.96.060407}. Applications of CV beams to solid state physics are, on the other hand, not explored well.

According to Ref.~\onlinecite{Youngworth:00}, the spatial profiles of the electric field $\bm{E}$ and the magnetic flux density $\bm{B}$ of the azimuthally polarized CV beam [Fig.~\ref{pol}$(c)$] propagating in a vacuum are, in the cylindrical coordinate $(\rho,\phi,z)$, given by
\begin{align} \label{fields}
E^\rho(\rho,\phi,z) &=E^z(\rho,\phi,z) = B^\phi(\rho,\phi,z) = 0 \\
E^\phi(\rho,\phi,z) &= 2 A \int^{\alpha}_{0}  \sin\theta f(\rho,\theta,z,1) d\theta, \nonumber \\
B^\rho(\rho,\phi,z) &= -\frac{2}{i c} A \int^{\alpha}_{0}  \sin\theta \cos\theta f(\rho,\theta,z,1) d\theta, \nonumber\\
B^z(\rho,\phi,z) &= \frac{2}{i c} A \int^{\alpha}_{0} \sin^2\theta f(\rho,\theta, z,0)d\theta  \nonumber\\
f(\rho,\theta,z, n) &= \cos^{\frac{1}{2}}(\theta)\ell_0 (\theta) J_n(k\rho \sin\theta) e^{i k z \cos\theta} \nonumber
\end{align}
where $c$ is the speed of light in a vacuum. The constant $\alpha$ specifies the size of the entrance pupil and $A$ gives the field amplitude. Here $J_n(x)$ is the Bessel function and $k$ is the wavenumber. We see that at $\rho = 0$ fields are vanishing except the $z$-component of magnetic field. The details of the apodization function $\ell_0 (\theta)$ and the choice of $\alpha$ depend on the pupil but the above property holds irrespective to its choice. 

\begin{figure}[htbp]
\centering
\includegraphics[width = 150mm]{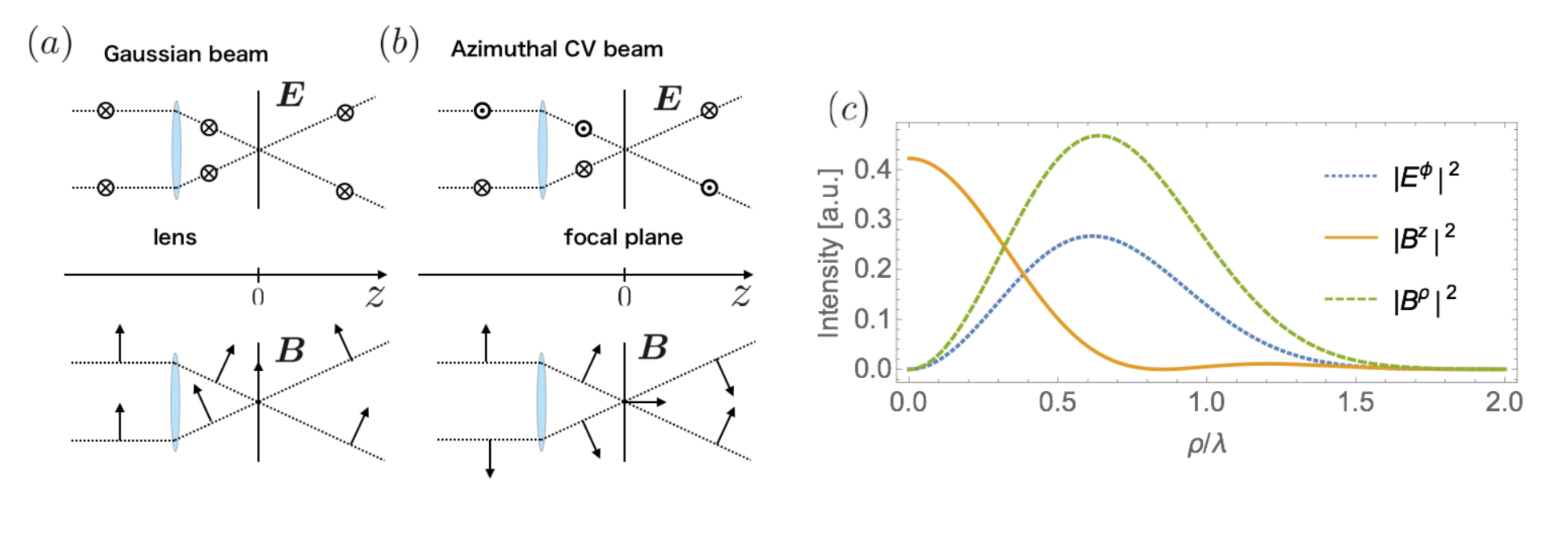}
       \caption{$(a)$, $(b)$: Focusing of the Gaussian beam and the azimuthal CV beam with a high numerical aperture lens. The characteristic spatial profile of the polarization vector of the azimuthal beam results in the vanishing electric field and the growth of the longitudinal magnetic field near the focus. $(c)$ Intensity distribution of the azimuthal beam in the focal plane at $z= 0$. The horizontal axis is the distance from the center of the focal plane. }
              \label{beam profile}
  \end{figure}

As a whole, the field distribution of the tightly focused azimuthally polarized CV beam is given as Fig.~\ref{beam profile}. Following Ref.~\onlinecite{Youngworth:00}, we take the pupil apodization function to be:
$\ell_0(\theta) = \exp \left[ - \beta^2 \left(\frac{\sin \theta}{\sin \alpha}\right)^2 \right ] J_1\left(2 \beta \frac{\sin\theta}{\sin\alpha}\right)$
and take $\beta$, the ratio of the radius of the pupil and the beam waist to be $1.5$. The parameter $\alpha$ is defined as $\alpha = \sin^{-1}({\rm NA}/n)$ where $n = 1.0$ is the refractive index of the vacuum and NA $ = 0.95$ is the numerical aperture of the lens. There exists a region where the ``longitudinal" magnetic field becomes dominant over the electric field (and other components of the magnetic field) [see Fig.~\ref{beam profile}~($b$), ($c$)]. Therefore, if a sample with its size well smaller than the wavelength is placed at the focus, we can virtually apply the desired ``pure magnetic field" for the nonequilibrium magnetic oscillation measurement [see Fig.~\ref{schem} $(b)$]. We stress that as the focusing becomes tighter, the longitudinal field does more prominent. The temporal profile of the electromagnetic fields is obtained just by multiplying $\exp(-i \omega t)$ to Eq.~\eqref{fields} and taking their real parts. The longitudinal part of the magnetic flux density $B^z$ oscillates in time.

\section{Magnetic oscillation of conducting magnets} 
We have discussed how the CV beam can be utilized to extend the magnetic oscillation measurement. In the following, taking a simple concrete model, we see how the magnetic oscillation signal depends on the configuration of localized magnetic moments. We consider the following square lattice tight-binding model under the CV beam whose electrons are coupled with localized moments through the exchange coupling. The Hamiltonian is given as
\begin{align}\label{model}
H = &-t\sum_{\braket{{\bm r}, {\bm r'}}, \sigma}c^{\dagger}_{{\bm r},\sigma}c_{{\bm r'},\sigma}e^{-i \frac{e}{\hbar}\int^{\bm r'}_{\bm r}\bm{A}(\bm{x})\cdot d\bm{x}} \nonumber \\& -2\mu_B B\sum_{{\bm r}}s^z_{\bm r}  - 2J_{ex}\sum_{{\bm r}, \alpha,\beta} \bm{m}_{{\bm r}}\cdot \bm{s}_{\bm r}.
\end{align}
The symbol $c_{\bm{r}, \sigma}$ is the annihilation operator of a conduction electron at site $\bm{r}$ with spin $\sigma$, and the dimensionless vector $\bm{m}_{\bm r}$ is the localized magnetic moment at site $\bm{r}$.
The first term describes the nearest-neighbor hopping with amplitude $t$ which is typically of the order of eV. The magnetic flux density $B$ of the beam is introduced by the Peierls substitution with the vector potential $\bm{A}(\bm{r}) = (By,0,0)$. The second term is the Zeeman coupling for the electron spins $(\bm{s}_{{\bm r}})_{\alpha,\beta} = c^{\dagger}_{{\bm r},\alpha}\bm{\sigma}_{\alpha,\beta}c_{{\bm r},\beta}$. Here the magnetic flux density is treated to be static because, as we discussed before, a pulse of the CV beam with sufficiently long duration works as a static one for conduction electrons. The third term is the exchange coupling with the coupling constant $J_{ex}$. Through the exchange coupling, conduction electrons feel the effective magnetic field in the direction of the localized magnetic moment $\bm{m}_{{\bm r}}$ at site $\bm r$. The exchange coupling $J_{ex}$ typically takes the value of the order of sub eV to eV. For example, in pyrochlore iridates R${}_{2}$Ir${}_{2}$O${}_{7}$, $J_{ex}$ is the exchange coupling between $f$ and $d$ orbitals and is considered to be around 5 $\%$ of the hopping $t$~\cite{PhysRevB.86.235129}.  If we regard the $J_{ex}$ as the Hund coupling in transition-metal-compounds like Mn oxides, its energy scale is eV~\cite{PhysRev.82.403,PhysRev.100.675}. Below we consider three different magnetic structures; ferromagnetic [$\bm{m}_{\bm r} = (0, 0, 1)$ for all ${\bm r}$], antiferromagnetic, and ferrimagnetic. For the antiferromagnetic and ferrimagnetic cases, we divide the system into two sublattices $A$ and $B$ and then define $\bm{m}_{{\bm r} \in A} = (0,0,1)$, $\bm{m}_{{\bm r} \in B} = (0,0,-1)$ for antiferromagnetic and $\bm{m}_{{\bm r} \in A} = (0,0,1)$, $m_{{\bm r} \in B} = (0,0,-0.5)$ for ferrimagnetic cases. In the following, we take $t = -3$, $J_{ex} = 2$ in the unit of electron volt and the lattice constant to be $a = 5$ \AA. We measure the magnetic flux density in the unit of Tesla. We again emphasize that we are assuming that the magnetic moments $\bm{m}_{\bm r}$ are independent on the laser magnetic flux density $B$ due to their timescale discrepancy.

The conducting-electron spin polarization of the model Eq.~\eqref{model} at zero temperature is given as:
\begin{align}
\braket{s^z_{\rm tot}}_B = \sum_{E_n < E_F} \braket{E_n| s^z_{\rm tot} | E_n},
\end{align}
where $s^z_{\rm tot}=\sum_{\bm r}s^z_{\bm r}$ is the $z$ component of the total conducting-electron spin, and $\ket{E_n}$ is the single-electron eigenstate with eigenenergy $E_n$. We denote the Fermi energy as $E_F$. Assuming the discrepancy in the relaxation timescale, we study this electron spin polarization as a function of the applied magnetic field of the CV beam.

\begin{figure}[htbp]
\centering
\includegraphics[width = 140mm]{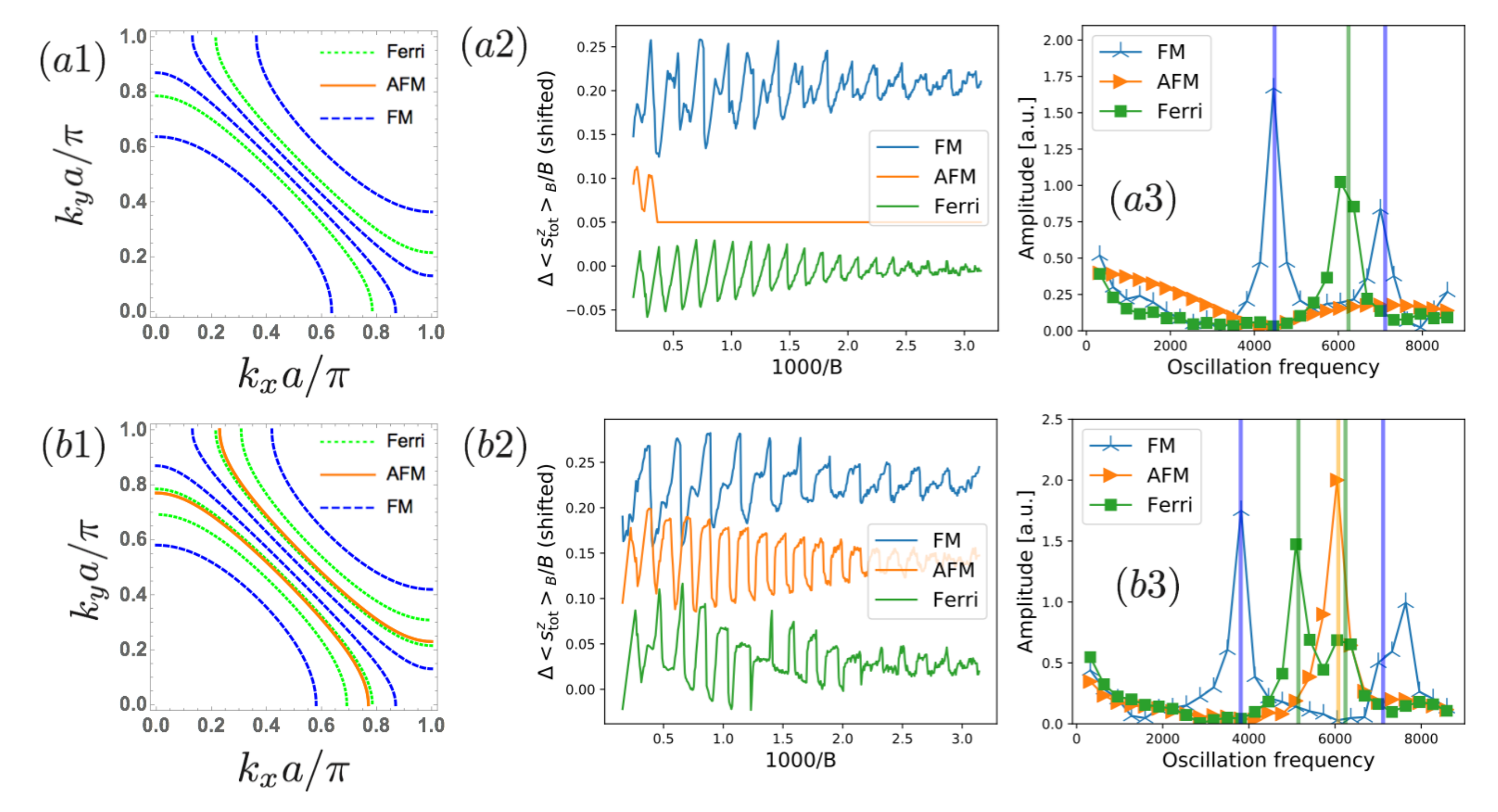}
       \caption{Calculation results of the magnetic oscillation of the model \eqref{model} on a lattice with linear dimensions $L_x = L_y = 30 a$ for $(a1$-$a3)$: $E_F = -1.5$ and $(b1$-$b3)$: $E_F = -2.5$. We consider three (ferromagnetic, antiferromagnetic, and ferrimagnetic) magnetic orders. $(a1,b1)$: Fermi surface structures for each magnetic order. $(a2, b2)$: field dependence of the electron spin polarization $\Delta\braket{s^z_{\rm tot}}_B = \braket{s^z_{\rm tot}}_B-\braket{s^z_{\rm tot}}_0$. The origins are shifted for visibility. These oscillations are Fourier transformed to obtain panels $(a3, b3)$. The vertical lines in these panels correspond to the oscillation frequency calculated from the area of each extremal Fermi surface [see Eq.~\eqref{freq_theory}].}
       \label{oscillation}
  \end{figure}

Figure~\ref{oscillation} summarizes the calculation on a square lattice with linear dimensions $L_x = L_y = 30 a$. Panels $(a1$-$a3)$ and $(b1$-$b3)$ are the results for the different values of the Fermi energy $E_F = -1.5$ and  $E_F = -2.5$ respectively. In Fig.~\ref{oscillation}~$(a1, b1)$, we present the Fermi surfaces for each magnetic structure. The magnetic oscillations of the electron spin polarization as a function of the inverse magnetic flux density are shown in Fig.~\ref{oscillation}$(a2, b2)$ and are Fourier transformed as shown in Fig.~\ref{oscillation}~$(a3, b3)$. In our setup with $t = -3$ and $J_{ex} = 2$, the band edge is at $E_F = -1$ for the ferrimagnetic and $E_F = -2$ for the antiferromagnetic cases. Therefore, as we see in Fig.~\ref{oscillation}$~(a1)$, there is no Fermi surface at $E_F = -1.5$ for the antiferromagnetic case. The oscillation frequencies indeed depend on the magnetic structure and are consistent with the predictions from the Fermi surface cross-section (vertical lines)~\cite{ashcroft1976solid,kittel2004introduction}:
\begin{align}\label{freq_theory}
\delta\left(\frac{1}{B}\right) =  \frac{2\pi e}{\hbar S_f}.
\end{align}
The left-hand side is the oscillation period of the magnetic susceptibility as a function of $1/B$ and $S_f$ is the area of the extremal orbit on the Fermi surface perpendicular to the applied magnetic field. Due to the small system size of our calculation, here we are taking an artificially strong magnetic field (remember that the Landau level degeneracy is determined by the total flux penetrating the entire system). In realistic situations, as is experimentally verified, magnetic flux density of O(1-10) Tesla would be sufficient enough to measure magnetic oscillations.
 
The result of Fig.~\ref{oscillation} clearly shows that the magnetic oscillation signal measured with the azimuthal CV beam would give the Fermi surface geometry reflecting the magnetic order. Thus, by measuring the spin polarization while changing the peak intensity and direction of the incident CV beam, we can read out the information of the full Fermi surface geometry in a magnetic-order-resolved way. If we use the magnetic oscillation in the conventional setup with the static magnetic field, we would obtain the signal in the ferromagnetic case irrespective to the actual magnetic order. 

\section{Discussion}
Here we discuss several important issues related to our proposal and calculation. First, we consider the effect of the exchange coupling on the dynamics of localized magnetic moments.
As we mentioned, the exchange coupling $J_{ex}$ is typically of the order of sub eV to eV. This seems to result in the fast dynamics of localized moments $\bm{m}_{\bm r}$ and break our assumption on the timescale discrepancy. The point is, though the exchange coupling itself has rather high energy, the laser-induced change in the electron spin polarization $\Delta\braket{\bm{s}_{\rm tot}}_B = \braket{\bm{s}_{\rm tot}}_B-\braket{\bm{s}_{\rm tot}}_0$ is very small [at most O($10^{-3}$) of the saturating value even for O($10$) Tesla as is read out from Fig.~\ref{oscillation} ($a2$, $b2$)]. The laser-induced dynamics of localized moments is determined by the effective magnetic field $J_{ex}\braket{\bm{s}_{\bm r}}$, not the exchange coupling $J_{ex}$ itself. Although $J_{ex}$ is large as it is,  the change in the local effective magnetic field is reduced by the small factor $\Delta\braket{\bm{s}_{\bm r}}_B =\braket{\bm{s}_{\bm r}}_B-\braket{\bm{s}_{\bm r}}_0$. When the initial state has no net spin polarization $\braket{\bm{s}_{\bm r}}_0 = 0$, the effective field determining the the dynamics of the localized moments is given by $J_{ex}\Delta\braket{\bm{s}_{\bm r}}_B$ and thus the dynamics turns out to be slow. On the other hand, when there is non-vanishing $ \braket{\bm{s}_{\bm{r}}}_0$, after the laser irradiation the spins $\bm{m}_{\bm r}$ start to precess around the effective magnetic field determined by $\braket{\bm{s}_{\bm{r}}}_B$. In this case, the period of the precession is indeed not slow. However, because $\Delta\braket{\bm{s}_{\bm{r}}}_B$ is very small, namely the directions of the effective magnetic fields before and after the application of the CV beam are almost identical, the amplitude of the magnetization precession is quite small. We can then virtually neglect the change in the spin structure. Therefore, even though the exchange coupling is strong, the laser-induced dynamics is slow, or the magnetization change can be ignored.

To observe the magnetic oscillation, typically we need a magnetic field pulse with the magnetic flux density of O(1-10) Tesla. As we mentioned, the frequency of the incident beam should be around THz or far-infrared. In the THz region, the magnetic field pulse of O(1) Tesla is just becoming possible, and it would take a while to reach O(10) Tesla. On the other hand, in the far-infrared region, it is much easier to achieve that.  For example, CO$_{2}$ lasers have been widely used for industrial purposes requiring the high intensity such as laser ablation or welding. Although CO$_{2}$ laser itself has the central wavelength of $10$ $\mu$m and its pulse duration may be too short (approximately 30 fs) for electrons to get relaxed, by using an array of half-cycle pulses we can synthesize a longer duration pulse. Moreover, we can use CO$_{2}$ lasers to pump other molecular lasers with longer wavelength such as CH$_{3}$OH lasers~\cite{Telles:1999aa}.

The effect of heating by the laser irradiation should also be considered. In the magnetic oscillation measurement, when the (electron) temperature becomes comparable with the bandwidth, the oscillation signal becomes vague. The primary source of the heating under irradiation of lasers is the coupling between the electric field and the electrons. Since the CV beam has a suppressed electric field near the focus, the heating is much weaker compared to that caused by the Gaussian beam. Nevertheless, there will exist a small change in the temperature. Therefore, applying our method to extremely heavy fermion systems with very small electron bandwidth could be problematic.

On top of the pump laser pulse, the way of measuring magnetic oscillations is also important for our setup. 
Since our proposal stands on the nonequilibrium property of the electron-spin coupled system, we have to measure the nonequilibrium spin polarization in the optical timescale. One way to achieve this is to use the magneto-optical effect such as Kerr or Faraday effects which is already a common way of measuring the magnetization. To measure the nonequilibrium magnetic oscillation, the temporal resolution of those magneto-optical measurements should be shorter than the CV pulse duration, thus we have to use the ultrashort femtosecond pulses to probe the spin density. We can also measure physical quantities other than the spin density. Because the magnetic oscillation originates from the Landau tube structure, there should appear oscillating behaviors in any physical quantities related to the electronic structure. For example, optical conductivity would be an option. 

In this paper, we have focused on the magnetic materials and their Fermi surfaces. However, the applicability of the proposed method is not limited to that. A notable example is the Fermi surface measurement of materials under ultra-high pressure. In producing ultra-high pressure stronger than GPa, commonly the diamond anvil cell~\cite{RevModPhys.55.65} is used, where the sample is placed in between two polished diamonds and pressed. In such a situation, performing the conventional de-Haas type measurement is experimentally difficult. Hence, Fermi surface of materials under ultra-high pressure is, even for non-magnetic metals, hard to be probed. On the other hand, since diamonds are highly transparent in the wide range of frequency including THz to far-infrared, our laser-based method is applicable there and will provide a new probe of electronic properties of generic metals under ultra-high pressure.

\section{conclusion}

In this paper, we propose an extension of the magnetic oscillation measurement for the Fermi surface geometry using the azimuthal cylindrical vector beams. In contrast to the conventional method using the static magnetic field, the proposed protocol allows us to apply that to conductors with localized magnetic moments with an arbitrary magnetic order. Hence, we can study the field-induced phase transitions in magnetic materials such as pyrochlore iridates, chiral magnets, and so on. The same method is applicable to the Fermi surface measurement of generic metals under ultra-high pressure where the conventional magnetic oscillation measurements are hard to be performed. 

The unique focusing property of the cylindrical beam would have various applications in condensed matter physics. If the sample size is sufficiently small as is assumed in this paper, the radial and the azimuthal cylindrical beams allow us to apply the electronic and magnetic fields independently. 
We can change their amplitude, relative angle or phase, and so on. 
This unique situation is an ideal playground to explore novel ways of controlling and measuring physical properties of matter in an ultrafast way.

\section{Acknowledgement}
We thank Atsushi Fujimori and Yasuhiro Tada for useful comments. We thank Keitaro Kuwahara for comments on the applicability to the ultra-high pressure experiments. H. F. is supported by Advanced Leading Graduate Course for Photon Science (ALPS) of Japan Society for the Promotion of Science (JSPS) and JSPS KAKENHI Grant-in-Aid for JSPS Fellows Grant No.~JP16J04752. M. S. was supported by Grant-in-Aid for Scientific Research on Innovative Area, ”Nano Spin Conversion Science” (Grant No.17H05174), and JSPS KAKENHI (Grant No. JP17K05513 and No. JP15H02117). A part of the computation in this work has been done using the facilities of the Supercomputer Center, the Institute for Solid State Physics, the University of Tokyo.  

\end{document}